\documentclass[a4paper,11pt]{article}

\usepackage{ragged2e} 
\usepackage{changepage}

\usepackage{siunitx} 
\usepackage{amsmath}
\usepackage{bm} 
\usepackage{amssymb}
\usepackage{wasysym}
\usepackage{xfrac} 

\usepackage{graphicx}
\usepackage{subcaption}
\usepackage{sidecap}
\usepackage{wrapfig}
\usepackage{caption}
\usepackage{verbatim}
\usepackage[english]{babel}

\usepackage{array}
\usepackage{adjustbox}
\usepackage{float}
\usepackage{enumerate}
\usepackage{multirow}
\usepackage{multicol}
\usepackage[table,xcdraw]{xcolor}
\usepackage[round]{natbib}

\usepackage{csvsimple}
\usepackage{bbm}
\usepackage{pdflscape}
\usepackage{breqn}
\usepackage{lscape}
\usepackage{amsfonts}
\usepackage{latexsym}
 \usepackage[titletoc]{appendix}  
\usepackage[colorlinks=true, citecolor= blue]{hyperref} 
\hypersetup{linkcolor=black} 
\usepackage[all]{hypcap} 

\usepackage{enumitem}

\usepackage{cleveref}

\usepackage{setspace}
\usepackage{relsize} 

\usepackage{authblk}

\usepackage{tabularx}

\newcommand{\shortsmile}[1]{\overset{\scalebox{0.6}[0.6]{$\smile$}}{#1}\mspace{-1mu}}

\title{\vspace{-1cm} \huge \textbf{Distributed lag non-linear models with Laplacian-P-splines for analysis of spatially structured time series}}
\date{}

\author[1]{Sara Rutten
\footnote{These authors contributed equally to this work.}\textsuperscript{,}\footnote{Corresponding author. {\textit{E-mail address}: sara.rutten@uhasselt.be}}\textsuperscript{,}}
\author[1,2]{Bryan Sumalinab \textsuperscript{\textasteriskcentered,}}
\author[1]{Oswaldo Gressani}
\author[1,3]{Thomas Neyens}
\author[1]{Elisa Duarte}
\author[1,4]{Niel Hens}
\author[1]{Christel Faes}

\affil[1]{Interuniversity Institute for Biostatistics and statistical Bioinformatics (I-BioStat), Data Science Institute (DSI), Hasselt University, Hasselt, Belgium}
\affil[2]{Department of Mathematics and Statistics, College of Science and Mathematics, Mindanao State University - Iligan Institute of Technology, Iligan City, Philippines}
\affil[3]{L-BioStat, Department of Public Health and Primary Care, KU Leuven, Leuven, Belgium}
\affil[4]{Centre for Health Economic Research and Modelling Infectious Diseases (CHERMID), Vaccine \& Infectious Disease Institute, Antwerp University, Antwerp, Belgium}

\begin{document}
\maketitle

\newpage
\begin{abstract}
  \noindent Distributed lag non-linear models (DLNM) have gained popularity for modeling nonlinear lagged relationships between exposures and outcomes. When applied to spatially referenced data, these models must account for spatial dependence, a challenge that has yet to be thoroughly explored within the penalized DLNM framework. This gap is mainly due to the complex model structure and high computational demands, particularly when dealing with large spatio-temporal datasets. To address this, we propose a novel Bayesian DLNM-Laplacian-P-splines (DLNM-LPS) approach that incorporates spatial dependence using conditional autoregressive (CAR) priors, a method commonly applied in disease mapping. Our approach offers a flexible framework for capturing nonlinear associations while accounting for spatial dependence. It uses the Laplace approximation to approximate the conditional posterior distribution of the regression parameters, eliminating the need for Markov chain Monte Carlo (MCMC) sampling, often used in Bayesian inference, thus improving computational efficiency. The methodology is evaluated through simulation studies and applied to analyze the relationship between temperature and mortality in London.\\
  
  
  \thispagestyle{empty}
\end{abstract}

\newpage
\clearpage\null

\pagenumbering{arabic}


\section{Introduction}

Distributed lag models (DLM), originally introduced within a time series setting, are used to model the influence of both current and past values of an explanatory variable on a response variable. They are also useful in epidemiology and public health, for example when examining the delayed effects of exposures such as temperature on health outcomes, showing that the impact of temperature on health is not always immediate. For instance, cold temperatures may have a prolonged impact, which means that low temperature could influence mortality rates for days or even weeks after the initial exposure \citep{analitis2008, Montero2010}. \cite{gasparrini2010} extended the DLM framework towards distributed lag non-linear models (DLNM). This extension offers greater flexibility in capturing nonlinear relationships between the explanatory and response variables. By using basis functions, DLNMs can model complex dependencies, allowing for a more flexible representation of how past values of an explanatory variable influence current health outcomes. This is helpful, as exposure-health relationships are rarely linear.  The DLNM framework has been further extended using a penalized spline approach to manage the flexibility of the fitted model \citep{gasparrini2017}. Penalized splines provide a method for smoothing the data in a flexible way without the risk of overfitting. This approach ensures that the model can adapt to the underlying data structure, capturing essential patterns. However, due to the high computational demands of this approach, it is rarely used in practice.\\



When data are collected across multiple time periods at a fine geographical level, aggregating them for an entire area or over large regions can introduce confounding and lead to biased results, especially when there is significant local variability. This variability can arise, for example, from differences in unmeasured characteristics unique to each area. Small-area analysis improves accuracy by keeping finer details but also increases data size and computational demands, requiring computationally efficient methods, especially for complex models like DLNM.  Additionally, Tobler's first law of geography implies that spatially close locations are likely to exhibit similar characteristics, including health outcomes, because they share similar values of regional factors. Therefore, dependence between the different areas is a realistic assumption, requiring a modeling strategy that accounts for the spatial structure of the data which may improve the model precision. In disease mapping, conditional autoregressive (CAR) models \citep{besag1974} are often used to analyze the spatial distribution of diseases. CAR-type models are statistical models, accounting for spatial autocorrelation and assuming that the value of a variable in a given area is conditionally dependent on the values in neighboring areas. In the DLNM framework, \cite{chien2018} account for spatial heterogeneity by adding a spatial random effect with a CAR prior. However, they noted the longer computational time required when incorporating the spatial random effect. \cite{lowe2021} account not only for spatial correlation but also for temporal correlation by introducing spatiotemporal random effects and leveraging integrated nested Laplace approximations for computational efficiency. \cite{gasparrini2022} accounts for area-specific characteristics but does not model these area-specific effects explicitly and instead uses a conditional Poisson model with a case-time series design. However, all of these proposed models are implemented using non-penalized DLNMs, thereby limiting flexibility in modeling the exposure-lag relationship, as results may be affected by the choice of the number of basis functions for the exposure and lag variables. The penalized version of the DLNM \citep{gasparrini2017} introduces an additional layer of complexity, which, when combined with spatial effects, needs a more complex and efficient modeling approach.  \\

We extend the penalized DLNM methododology of \cite{gasparrini2017} by proposing a fast Bayesian methodology based on Laplacian-P-splines (DLNM-LPS) to account for extra spatial variability and dependency. In the proposed model, P-splines \citep{eilers1996, eilers2015twenty} are used to capture the non-linear relationships between exposure and health, as well as the lagged effects. The penalties are effortlessly constructed and translated into priors for the B-spline coefficients within the Bayesian framework \citep{lang2004}. Furthermore, by unifying Laplace approximations and P-splines, we propose a flexible framework with significant computational advantages, particularly for large datasets. The Laplacian-P-splines (LPS) approach was originally developed by \cite{gressani2018} in the context of cure survival models and later extended for generalized additive models \citep{gressani2021}. The Laplace approximation is based on a second-order Taylor expansion around the mode of the target posterior distribution and yields a Gaussian approximation to the target. Compared to Markov chain Monte Carlo (MCMC) methods, which are commonly used in Bayesian inference but can be computationally intensive and exhibit slow convergence, the Laplace approximation reduces the computational burden and enhances the speed of the inferential process, making it feasible to handle large datasets. This efficiency is particularly valuable when large amounts of spatial and temporal data are involved. The spatially structured random effects in the proposed model mainly account for unmeasured spatial heterogeneity and spatial dependencies using CAR priors.\\ 

Section 2 reviews the DLNM and explains how the penalized DLNM can be embedded into a Bayesian framework. Then, we provide derivations for approximating the posterior distribution of the main parameters, leading to fast inference. Performance of the proposed approach is assessed through simulation studies in Section 3 and the methodology is applied in Section 4 to a dataset containing deaths and temperature readings for the years 2006 and 2013.  

\section{Methodology}

\subsection*{Distributed lag non-linear model (DLNM)}
Suppose that we have a time series of (mortality) counts $y_{tj}$ at time $t=1,\dots, T$ for each spatial location $j=1,\dots,J$. Furthermore, the presence of a predictor $x$ with delayed effects on the response $y_{tj}$ is assumed. Following the DLNM formulation of \cite{gasparrini2010}, this predictor can be included through a smoothing function $s$, defined as:
\begin{align}
\label{eqn:DLNM}
    s(x_{t,j}, \ldots , x_{t-L,j} ; \theta_{ik})&=\sum_{i=1}^{v_x} \sum_{k=1}^{v_l}\left(\sum_{l=0}^L\widetilde{b}_i(x_{t-l,j})\shortsmile{b}_k(l) \right)\theta_{ik},
\end{align}
where $\{\widetilde{b}_i\}_{i=1}^{v_x}$ and $\{\shortsmile{b}_k\}_{k=1}^{v_l}$ are B-spline basis functions evaluated on the predictor ($x_{t-l,j}$) and lag variable ($l$), respectively. Let $\boldsymbol{\ell} = (0, 1, \ldots, L)^{\top}$ denote the vector of lagged values and $\mathbf{q}_{t,j} = (x_{t,j} , x_{t-1,j}, \ldots , x_{t-L,j})^{\top}$ the vector of lagged occurrence of a covariate $x_{t-l,j}$ for lag $l$ at each time $t$ and location $j$. Applying the basis functions $\widetilde{b} (\cdot)$ and $\shortsmile{b}(\cdot)$ to the vectors $\mathbf{q}_{t,j}$ and $\boldsymbol{\ell}$ will yield basis matrices $\widetilde{B}_{t,j}$ and $\shortsmile{B}$ with dimensions $(L+1) \times v_x$ and $(L+1) \times v_l$, respectively. \cite{gasparrini2017} showed that equation \eqref{eqn:DLNM} can also be written in matrix form as: 
\begin{align*}
    s(x_{t,j}, \ldots , x_{t-L,j};\boldsymbol{\theta})&= \left( \mathbf{1}^{\top}_{L + 1} \, A_{t,j} \right) \boldsymbol{\theta}= \boldsymbol{w}_{tj}^{\top}\boldsymbol{\theta}, 
\end{align*}
with $\mathbf{1}$ a vector of ones of appropriate dimension and $\boldsymbol{w}_{tj}$ with dimension $1 \times (v_x \times v_l)$ is derived from the matrix $A_{t,j} = \left( \widetilde{B}_{t,j} \otimes 1_{v_l}^\top \right) \odot \left( 1_{v_x}^\top \otimes \shortsmile{B} \right)$ where the symbols $\otimes$ and $\odot$ denotes the Kronecker and Hadamard products, respectively. The parameter vector $\boldsymbol{\theta}$ contains all the $v_x \times v_l$ coefficients given by $\boldsymbol{\theta} = (\theta_{11},\dots, \theta_{1v_{l}},\theta_{21},\dots, \theta_{2v_{l}},\dots \theta_{v_{x}1},\dots, \theta_{v_{x}v_{l}})^{\top}$ and  the full cross basis matrix $W$ with dimension $TJ \times (v_x \times v_l)$ is obtained by deriving $w_{tj}$ for all $t=1\dots,T$ and $j=1,\dots,J$. A more detailed description of DLNM can be found in the paper of \cite{gasparrini2010, gasparrini2017}. \\


Assuming a Poisson distribution on $y_{tj}$ with mean $\mathbb{E}(y_{tj}) = \mu_{tj}$, we have the following model:
\[\log(\mu_{tj}) = \boldsymbol{z}^{\top}_{tj}\boldsymbol{\beta} +  \boldsymbol{w}^{\top}_{tj}\boldsymbol{\theta} + u_j,\]

where $\boldsymbol{z}_{tj}$ is the \textit{tj}th row of the design matrix of $p$ linear covariates (including the intercept) with dimension $1 \times (p+1)$ and associated  coefficient $\boldsymbol{\beta}$, $\boldsymbol{w}_{tj}$ and $\boldsymbol{\theta}$ as defined previously, and $u_j$ is a spatial location-specific random intercept. Let $\boldsymbol{y} = \{y_{tj}: t=1,\dots,T; j=1,\dots,J\}$ denote the vector of mortality counts with the mean vector $\mathbb{E}(\boldsymbol{y}) = \boldsymbol{\mu}$. In matrix form, the model can then be written as:
\[\log(\boldsymbol{\mu}) = Z\boldsymbol{\beta} + W\boldsymbol{\theta} + M\boldsymbol{u} = H\boldsymbol{\xi},\]
where $H=[Z:W:M]$, $\boldsymbol{\xi}=(\boldsymbol{\beta}^{\top},\boldsymbol{\theta}^{\top}, \boldsymbol{u}^{\top})^{\top}$, $Z$ denotes the design matrix of the $p$ linear covariates (including the intercept) with dimension $TJ \times (p+1)$, and $M$ is the design matrix for the random vector $\boldsymbol{u} = (u_1, u_2, \dots , u_J)^{\top}$. \\

A topic often of interest in epidemiological analysis is the attributable fraction (AF) i.e.\ the fraction of outcome cases attributed to the exposure $x$. \cite{Gasparrini2014} developed this concept in the context of DLNMs. They introduced two different time perspectives: a forward perspective and backward perspective. Following these two perspectives, the forward and backward AF at time $t$, compared to a reference exposure $x_0$, can be defined by
\begin{align*}
    AF_{f}(x,t) &= 1-\exp \biggl\{-\sum_{i=1}^{v_x}\sum_{k=1}^{v_l}\biggl(\sum_{l=0}^{L}{\bigl(\widetilde{b}_i(x_{t})-\widetilde{b}_i(x_0)\bigl)\shortsmile{b}_k(l)\biggl)\theta_{ik}}\biggl\} 
\end{align*}
    and
\begin{align*}
    AF_{b}(x,t) &= 1-\exp\biggl\{ -\sum_{i=1}^{v_x}\sum_{k=1}^{v_l}\biggl(\sum_{l=0}^{L}{\bigl(\widetilde{b}_i(x_{t-l})-\widetilde{b}_i(x_0)\bigl)\shortsmile{b}_k(l)\biggl)\theta_{ik}})\biggl\} 
\end{align*}
respectively. Hence, while the forward perspective specifies a future risk resulting from a current exposure, the backward perspective quantifies the current risk resulting from past exposures. \\

\subsection*{Spatial random effects}
Different spatial random effect models can be assumed for the random effects $u_j$. The simplest choice for $\boldsymbol{u}$ is to assume $u_j \stackrel{iid}{\sim} \mathcal{N}(0, \tau^{-1})$ for $j=1,\dots,J$. However, to account for spatial correlation, the most commonly used prior for $u_j$  is based on CAR models used in disease mapping (see, e.g., \cite{lee2011} for a review of these models). The first model that is considered here uses the intrinsic autoregressive (ICAR) model proposed by \cite{besag1991} of the form $(u_j|u_h, j \neq h) \sim \mathcal{N}\left(n_j^{-1} \sum\limits_{j \sim h}u_h, (\tau n_j)^{-1}\right)$ where $j \sim h$ denotes that locations $j$ and $h$ are neighbours and $n_j$ is the number of neighbours. This conditional prior can also be translated into a joint prior for the random vector $\boldsymbol{u}$ \citep{fahrmeir2001} such that:

\begin{equation}
    \label{eqn:ICAR} p(\boldsymbol{u}|\tau, \Lambda ) \propto  \exp\left(-\frac{\tau}{2}\boldsymbol{u}^{\top} \Lambda \boldsymbol{u}\right),
\end{equation}

where $\Lambda $ has elements

\[ r_{jh} = \begin{cases} 
      n_j & j = h \\
      -1 & j \sim h \\
      0 & \text{otherwise}. 
   \end{cases}
\]
Although the covariance matrix $\Lambda $ is usually rank-deficient and thus not invertible, this does not cause numerical problems since inversion of matrix~$\Lambda$ is not required by our methodology. The prior specified in \eqref{eqn:ICAR} may not be appropriate if the data is weakly spatially correlated. Hence, an extension of the ICAR prior also proposed by \cite{besag1991} is to add independent random effects such that $u_j = u_{1j} + u_{2j}$, where $u_{1j} \sim \mathcal{N}(0, \tau_1)$ and $u_{2j}$ is the CAR prior previously described in \eqref{eqn:ICAR}. This is also called the convolution or BYM model. Finally, another model that is considered is the Leroux prior \citep{leroux1999} that assumes a multivariate Gaussian prior for $\boldsymbol{u}$ given by:
\begin{equation*}
    (\boldsymbol{u}|\tau, \rho, \Lambda ) \sim \mathcal{N}_{J}\left(0, \left[\tau (\rho \Lambda  + (1-\rho)I_J)\right]^{-1}\right),
\end{equation*}
where $0 \leq \rho < 1$, with $\rho$ values close to zero representing a weak spatial correlation while values close to 1 correspond to a strong spatial correlation.

\subsection*{Bayesian model formulation}
We embed the spatially-explicit DLNM with P-splines in the Bayesian inference framework. This has the advantage of providing an automatic estimation scheme for the smoothing parameters. The difference penalties can be easily incorporated into a Bayesian framework by using random walk priors. Moreover, the Bayesian methodology naturally quantifies uncertainties, potentially providing more accurate estimates of variability and improved coverage. Additionally, quantities like exceedance probabilities are easily obtained. The Bayesian framework also enables the straightforward inclusion of spatial dependence through the widely used CAR priors for spatial random effects in disease mapping. \\

We use Bayesian P-splines to smooth both the exposure and lag variable \citep{lang2004}, which counterbalance the associated flexibility by imposing a discrete roughness penalty on
contiguous B-spline coefficients.  Let $D^m_{v_x}$ and $D^m_{v_l}$ denote the \textit{m}th order difference matrix with dimensions $(v_x - m) \times v_x$ and $(v_l - m) \times v_l$, respectively. In this paper, a second order difference penalty is used ($m=2$). The second order difference matrix is of the form:

 \[
D^2 = \begin{pmatrix}
1 & -2 & 1 & 0 & \cdots & 0 & 0 & 0 \\
0 & 1 & -2 & 1& \cdots & 0 & 0 & 0 \\
\vdots & \vdots & \vdots & \vdots & \ddots & \vdots & \vdots & \vdots\\
0 & 0 & 0 & 0 & \cdots & 1 & -2 & 1
\end{pmatrix}.
\]

\vspace{0.2cm}

 For ease of notation, let $D_{v_x}=D^m_{v_x}$ and $D_{v_l}=D^m_{v_l}$. Define the penalty matrices $S_x=D_{v_x}^{\top}D_{v_x} + \delta I_{v_x}$ and  $S_{l}=D_{v_l}^{\top}D_{v_l} + \delta I_{v_l}$, where $\delta$ is a small number (say $\delta=10^{-12}$), to ensure that the penalty matrices are full rank.  The penalty can be translated into a Gaussian prior for $\boldsymbol{\theta}$ \citep{lang2004} such that $(\boldsymbol{\theta}|\boldsymbol{\lambda}) \sim \mathcal{N}_{dim(\boldsymbol{\theta})}(\boldsymbol{0},\mathcal{P}^{-1}(\boldsymbol{\lambda}))$, where $\boldsymbol{\lambda}=(\lambda_x,\lambda_l)^{\top}$ is the penalty vector and $\mathcal{P}(\boldsymbol{\lambda})=\lambda_x (S_x\otimes I_{v_l}) + \lambda_l (I_{v_x} \otimes S_l)$. Additional penalties can also be imposed for the lag variable, for example, forcing the lag effect to approach zero for higher delays \citep{gasparrini2017}. 
 
 A Gaussian prior is assumed for the fixed effect parameters $\boldsymbol{\beta}$, namely $\boldsymbol{\beta} \sim \mathcal{N}_{dim(\boldsymbol{\beta})}(\boldsymbol{0},{V}^{-1}_{\boldsymbol{\beta}})$ with ${V}_{\boldsymbol{\beta}} = \zeta {I}_{p+1}$ ($\zeta=10^{-5}$). 
 Furthermore, denote the precision matrix for the spatial random effect $\boldsymbol{u}$ by $G$, that is, $\text{Cov}(\boldsymbol{u}|\tau, \tau_1, \tau_2, \rho, \Lambda) = G^{-1}$. The precision for $\boldsymbol{\xi}=(\boldsymbol{\beta}^{\top},\boldsymbol{\theta}^{\top}, \boldsymbol{u}^{\top})^{\top}$ is then given by $Q = \text{blkdiag}(V_{\boldsymbol{\beta}}, \mathcal{P}(\boldsymbol{\lambda}), G)$ where $\text{blkdiag}(\cdot)$ denotes a block diagonal matrix. 
 
 Finally, the following priors are assumed for the penalty parameters: $ (\lambda_x|\delta_x) \sim \mathcal{G}\left(\frac{\nu}{2},\frac{\nu \delta_x}{2}\right)$, $
 (\lambda_l|\delta_l) \sim \mathcal{G}\left(\frac{\nu}{2},\frac{\nu \delta_l}{2}\right)$  and $\delta_x \sim \mathcal{G}(a_{x},b_{x})$, and $
\delta_l \sim \mathcal{G}(a_{l},b_{l})$ where  $\mathcal{G}(a,b)$ denotes a Gamma distribution with mean $a/b$ and variance $a/b^2$. This robust prior specification for the penalty parameters is based on \cite{jullion2007} where $a=b$ is chosen to be sufficiently small (e.g., $10^{-5}$) and $\nu$ is fixed ($\nu=3$). These priors are also adapted for the precision parameters of the random effects.

The Bayesian model is then summarized as follows:
\begin{align*}
    &\begin{array}{ll}
        (\boldsymbol{y}|\boldsymbol{\xi}) \sim \text{Poisson}(\boldsymbol{\mu}) \text{ with } \log(\boldsymbol{\mu}) = H\boldsymbol{\xi}, & (\tau|\delta_\tau) \sim \mathcal{G}\left(\frac{\nu}{2},\frac{\nu \delta_{\tau}}{2}\right), \\
        (\boldsymbol{\xi}|\boldsymbol{\lambda}, \tau, \tau_1, \tau_2, \rho) \sim \mathcal{N}(\boldsymbol{0},Q^{-1}), & (\tau_1|\delta_{\tau_1}) \sim \mathcal{G}\left(\frac{\nu}{2},\frac{\nu \delta_{\tau_1}}{2}\right), \\
        (\lambda_x|\delta_x) \sim \mathcal{G}\left(\frac{\nu}{2},\frac{\nu \delta_x}{2}\right), & (\tau_2|\delta_{\tau_2}) \sim \mathcal{G}\left(\frac{\nu}{2},\frac{\nu \delta_{\tau_2}}{2}\right), \\
        (\lambda_l|\delta_l) \sim \mathcal{G}\left(\frac{\nu}{2},\frac{\nu \delta_l}{2}\right), & \delta_{\tau} \sim \mathcal{G}(a_{\tau},b_{\tau}), \\
        \delta_x \sim \mathcal{G}(a_{x},b_{x}), & \delta_{\tau_1} \sim \mathcal{G}(a_{\tau_1},b_{\tau_1}), \\
        \delta_l \sim \mathcal{G}(a_{l},b_{l}), & \delta_{\tau_2} \sim \mathcal{G}(a_{\tau_2},b_{\tau_2}), \\
        & \rho \sim \text{Beta}\left(\frac{1}{2}, \frac{1}{2}\right).
    \end{array}
\end{align*}
It is important to note that the inclusion of the parameters $\tau$, $\tau_1$, $\tau_2$, and $\rho$ in the model depends on the prior assumptions for the spatial random effects.

\subsection*{Laplace approximation to the conditional posterior for $\boldsymbol{\xi}$}
The LPS method is proposed to estimate the model parameters $\boldsymbol{\xi}$. 
In the first step, the Laplace approximation is derived of the conditional posterior of $\boldsymbol{\xi}$ given the penalty and random effect hyperparameters. This involves deriving the analytic gradient and Hessian of the conditional posterior with respect to $\boldsymbol{\xi}$. These are used as inputs in a Newton-Raphson algorithm, resulting in a Gaussian approximation to the joint conditional posterior distribution of~$\boldsymbol{\xi}$. The set of random effect hyperparameters is denoted as $\boldsymbol{\omega} = (\tau, \tau_1, \tau_2, \rho)^\top$. Using Bayes' theorem, the conditional posterior of $\boldsymbol{\xi}$ can be written as:
\begin{align*}
    p(\boldsymbol{\xi}| \boldsymbol{\lambda}, \boldsymbol{\omega}, \mathcal{D}) & \propto \mathcal{L}(\boldsymbol{\xi}, \boldsymbol{\omega}; \mathcal{D}) p(\boldsymbol{\xi}|\boldsymbol{\lambda}, \boldsymbol{\omega})\\
    & \propto \exp\bigg(\sum_{tj}\Big(y_{tj}\log(\mu_{tj})-\mu_{tj}\Big)\bigg) \exp\bigg( - \frac{1}{2}{\boldsymbol{\xi}}^{\top} Q \boldsymbol{\xi}\bigg),
\end{align*}
where $\mathcal{L}(\boldsymbol{\xi}, \boldsymbol{\omega}; \mathcal{D})$ denotes the likelihood function and $\mathcal{D}$ denotes the observed data. The log-conditional posterior of $\boldsymbol{\xi}$ is then given by:
\begin{equation}
\label{eqn:logposterior}
    \log  p(\boldsymbol{\xi}| \boldsymbol{\lambda}, \boldsymbol{\omega}, \mathcal{D}) \dot{=} \sum_{tj}\Big(y_{tj}\log(\mu_{tj})-\mu_{tj}\Big) - \frac{1}{2}{\boldsymbol{\xi}}^{\top} Q \boldsymbol{\xi},
\end{equation}
where $\dot{=}$ denotes equality up to an additive constant. From the theory of generalized linear models, it is known that the gradient and Hessian following from a Poisson log-likelihood are given by $\nabla_{\boldsymbol{\xi}}\log \mathcal{L}(\boldsymbol{\xi}, \boldsymbol{\omega}; \mathcal{D})=H^{\top}(\boldsymbol{y}-\boldsymbol{\mu})$ and  $\nabla^2_{\boldsymbol{\xi}}\log \mathcal{L}(\boldsymbol{\xi}, \boldsymbol{\omega}; \mathcal{D})=-H^{\top}VH$, respectively, where $V$ is a diagonal matrix with entries on the diagonal given by $\mu_{tj}$. Therefore, the gradient and Hessian for the log-conditional posterior of $\boldsymbol{\xi}$ are $\nabla_{\boldsymbol{\xi}}\log p(\boldsymbol{\xi}| \boldsymbol{\lambda}, \boldsymbol{\omega}, \mathcal{D}) = H^{\top}(\boldsymbol{y}-\boldsymbol{\mu})-Q {\boldsymbol{\xi}}$ and $ \nabla^2_{\boldsymbol{\xi}}\log p(\boldsymbol{\xi}| \boldsymbol{\lambda}, \boldsymbol{\omega}, \mathcal{D})=-(H^{\top}VH + Q).$ To obtain the mode of this log-conditional posterior, a Newton-Raphson algorithm is used. This results in the Laplace approximation of the conditional posterior of $\boldsymbol{\xi}$ being a multivariate Gaussian density denoted as $\widetilde{p}_G(\boldsymbol{\xi}| \boldsymbol{\lambda}, \boldsymbol{\omega}, \mathcal{D})=\mathcal{N}(\widehat{\boldsymbol{\xi}},\widehat{\Sigma})$, where $\widehat{\boldsymbol{\xi}}$ is the mode and $\widehat{\Sigma}$ is the variance-covariance obtained as the negative inverse of the Hessian matrix evaluated at the mode.

\subsection*{Hyperparameter optimization}

In the second step, we derive the (approximate) posterior distribution of the hyperparameters belonging to the penalization part, i.e.\ $\boldsymbol{\lambda} = (\lambda_x, \lambda_l)^{\top}$ and $\boldsymbol{\delta}_{\boldsymbol{\lambda}} = (\delta_x, \delta_l)^{\top}$, and the hyperparameters related to the random effects $\boldsymbol{\omega} = (\tau, \tau_1, \tau_2, \rho)^\top$ and $\boldsymbol{\delta_{\omega}} = (\delta_{\tau}, \delta_{\tau_1}, \delta_{\tau_2})^\top$. Using Bayes' theorem, the joint marginal posterior of the penalty and random effect parameters is: $$p(\boldsymbol{\lambda}, \boldsymbol{\delta}_{\boldsymbol{\lambda}}, \boldsymbol{\omega}, \boldsymbol{\delta_{\omega}}|\mathcal{D}) \propto \frac{\mathcal{L}(\boldsymbol{\xi}, \boldsymbol{\omega};\mathcal{D})p(\boldsymbol{\xi}|\boldsymbol{\lambda},\boldsymbol{\omega})p(\boldsymbol{\lambda}|\boldsymbol{\delta}_{\boldsymbol{\lambda}})p(\boldsymbol{\delta}_{\boldsymbol{\lambda}})p(\boldsymbol{\omega}|\boldsymbol{\delta_{\omega}})p(\boldsymbol{\delta_{\omega}})}{p(\boldsymbol{\xi}| \boldsymbol{\lambda}, \boldsymbol{\omega}, \mathcal{D})}.$$ Following \cite{rue2009}, this joint posterior can be approximated by replacing $p(\boldsymbol{\xi}| \boldsymbol{\lambda}, \boldsymbol{\omega}, \mathcal{D})$ with its Gaussian approximation $\widetilde{p}_G(\boldsymbol{\xi}| \boldsymbol{\lambda}, \boldsymbol{\omega}, \mathcal{D})$,  and by evaluating $\boldsymbol{\xi}$ at $\widehat{\boldsymbol{\xi}}$. Denote the estimate for $\mu_{tj}$ as $\widehat{\mu}_{tj}$, such that $\log(\widehat{\mu}_{tj}) = \textbf{h}_{tj}^{\top}\widehat{\boldsymbol{\xi}}$, where $\textbf{h}_{tj}^{\top}$
corresponds to the \textit{tj}th row of the design matrix $H$. Note that the determinant $|Q|^{\frac{1}{2}}$ in $ p({\boldsymbol{\xi}}|\boldsymbol{\lambda}, \boldsymbol{\omega})$ is $|Q|^{\frac{1}{2}}\propto |\mathcal{P}(\boldsymbol{\lambda})|^{\frac{1}{2}} \times |G|^{\frac{1}{2}}$ and $\widetilde{p}_G(\boldsymbol{\xi}| \boldsymbol{\lambda}, \boldsymbol{\omega}, \mathcal{D})\bigr|_{\boldsymbol{\xi}=\widehat{\boldsymbol{\xi}}} \propto |\widehat{\Sigma}|^{-\frac{1}{2}}$. Hence, the approximated joint marginal posterior of the hyperparameters is:

\begin{align*}
    \widetilde{p}(\boldsymbol{\lambda}, \boldsymbol{\delta}_{\boldsymbol{\lambda}}, \boldsymbol{\omega}, \boldsymbol{\delta_{\omega}}|\mathcal{D}) & \propto \frac{\mathcal{L}(\boldsymbol{\xi}, \boldsymbol{\omega};\mathcal{D})p(\boldsymbol{\xi}|\boldsymbol{\lambda},\boldsymbol{\omega})p(\boldsymbol{\lambda}|\boldsymbol{\delta}_{\boldsymbol{\lambda}})p(\boldsymbol{\delta}_{\boldsymbol{\lambda}})p(\boldsymbol{\omega}|\boldsymbol{\delta_{\omega}})p(\boldsymbol{\delta_{\omega}})}{\widetilde{p}_G(\boldsymbol{\xi}| \boldsymbol{\lambda}, \boldsymbol{\omega}, \mathcal{D})}\Bigr|_{\boldsymbol{\xi}=\widehat{\boldsymbol{\xi}}}\\
    & \propto \exp\bigg(\sum_{tj}\Big(y_{tj}\log(\widehat{\mu}_{tj})-\widehat{\mu}_{tj}\Big)\bigg) \times |\mathcal{P}(\boldsymbol{\lambda})|^{\frac{1}{2}} \exp\bigg( - \frac{1}{2}{\widehat{\boldsymbol{\xi}}}^{\top} Q \widehat{\boldsymbol{\xi}}\bigg)\\
    & \quad \times  \left(\lambda_x\right)^{(\frac{\nu}{2}-1)} \delta_x^{(\frac{\nu}{2}+a_x-1)} \exp(-\delta_x(b_x+\frac{\nu}{2}\lambda_x))\\
    & \quad \times \left(\lambda_l\right)^{(\frac{\nu}{2}-1)}  \delta_l^{(\frac{\nu}{2}+a_l-1)} \exp(-\delta_l(b_l+\frac{\nu}{2}\lambda_l)) \\
    & \quad \times |\widehat{\Sigma}|^{\frac{1}{2}}|G|^{\frac{1}{2}} p(\boldsymbol{\omega}|\boldsymbol{\delta_{\omega}})p(\boldsymbol{\delta_{\omega}}),
\end{align*}
where the term $p(\boldsymbol{\omega}|\boldsymbol{\delta_{\omega}})p(\boldsymbol{\delta_{\omega}})$ depends on the prior assumptions for the random effects. The hyperparameters $\boldsymbol{\delta}_{\boldsymbol{\lambda}}$ and $\boldsymbol{\delta}_{\boldsymbol{\omega}}$ can be integrated out from $p(\boldsymbol{\lambda}, \boldsymbol{\delta}_{\boldsymbol{\lambda}}, \boldsymbol{\omega}, \boldsymbol{\delta_{\omega}}|\mathcal{D})$ to obtain the marginal posterior of $\boldsymbol{\lambda}$ and $\boldsymbol{\omega}$ given by:
\begin{align*}
    \widetilde{p}(\boldsymbol{\lambda}, \boldsymbol{\omega}|\mathcal{D}) & = \int_0^\infty \cdots \int_0^\infty p(\boldsymbol{\lambda}, \boldsymbol{\delta}_{\boldsymbol{\lambda}}, \boldsymbol{\omega}, \boldsymbol{\delta}_{\boldsymbol{_{\omega}}}|\mathcal{D}) d\boldsymbol{\delta}_{\boldsymbol{\lambda}} \, d\boldsymbol{\delta}_{\boldsymbol{_{\omega}}}\\
    & \propto \exp\bigg(\sum_{tj}\Big(y_{tj}\log(\widehat{\mu}_{tj})-\widehat{\mu}_{tj}\Big)\bigg) \times |\mathcal{P}(\boldsymbol{\lambda})|^{\frac{1}{2}} \exp\bigg( - \frac{1}{2}{\widehat{\boldsymbol{\xi}}}^{\top} Q \widehat{\boldsymbol{\xi}}\bigg)\\
    & \quad \times \left(\lambda_x \lambda_l\right)^{(\frac{\nu}{2}-1)} \left(b_x+\frac{\nu}{2}\lambda_x\right)^{-(\frac{v}{2}+a_x)} \left(b_l+\frac{\nu}{2}\lambda_l\right)^{-(\frac{v}{2}+a_l)} |\widehat{\Sigma}|^{\frac{1}{2}}|G|^{\frac{1}{2}} p(\boldsymbol{\omega}).
\end{align*}

To ensure numerical stability, we log-transform the penalty vector $\boldsymbol{v}_{\boldsymbol{\lambda}}=(v_x,v_l)^{\top}=(\log(\lambda_x),\log(\lambda_l))^{\top}$. For the random effects, the precision parameters are log-transformed and logit transformation is used for the correlation parameter $\rho$ of the Leroux model since $0 \leq \rho < 1$. We denote this by $\boldsymbol{v}_{\boldsymbol{\omega}} = (v_{\tau}, v_{\tau_1}, v_{\tau_2}, v_{\rho})^\top = (\log(\tau), \log(\tau_1), \log(\tau_2), \log(\frac{\rho}{1-\rho}))^\top$. Note that the Jacobian of the transformation with respect to $\boldsymbol{v}_{\boldsymbol{\lambda}}$ is $\exp(v_x)\exp(v_l)$. The joint log-posterior of $\boldsymbol{v}_{\boldsymbol{\lambda}}$ and $\boldsymbol{v}_{\boldsymbol{\omega}}$  is then given by:

\begin{align}
     \log \widetilde{p}(\boldsymbol{v_{\lambda}}, \boldsymbol{v_{\omega}};\mathcal{D}) & \dot{=} \sum_{tj}\Big(y_{tj}\log(\widehat{\mu}_{tj})-\widehat{\mu}_{tj}\Big) + \frac{1}{2}\log |P(\boldsymbol{v}_{\boldsymbol{\lambda}})| - \frac{1}{2}{\widehat{\boldsymbol{\xi}}}^{\top} Q \widehat{\boldsymbol{\xi}} + {\frac{1}{2}} \log |\widehat{\Sigma}| \notag \\
     & \quad + \frac{\nu}{2}\left(v_x + v_l\right)- \left(\frac{\nu}{2}+a_x\right) \left(\log(b_x+\frac{\nu}{2}\exp(v_x))\right) \notag \\
     & \quad - \left(\frac{\nu}{2}+a_l\right) \left(\log(b_\delta+\frac{\nu}{2}\exp(v_t))\right) + \text{[\textit{random effect parameters}]} \label{eqn:log-posterior_hyper}
\end{align}

where the term `\textit{random effect parameters}' captures the contribution of the parameters related to the random effects given as follows:
\[
\begin{aligned}
    & \text{Independent and ICAR prior}: \quad \left(\frac{\nu + J}{2}\right) v_{\tau} - \left( \frac{\nu}{2} + a_{\tau} \right) \log \left( b_{\tau} + \frac{\nu}{2} \exp(v_{\tau}) \right), \\
    & \text{Convolution model}: \quad \left(\frac{\nu + J}{2}\right) v_{\tau_1} - \left( \frac{\nu}{2} + a_{\tau_1} \right) \log \left( b_{\tau_1} + \frac{\nu}{2} \exp(v_{\tau_1}) \right) \\
    &\hspace{10em} + \left(\frac{\nu + J}{2}\right) v_{\tau_2} - \left( \frac{\nu}{2} + a_{\tau_2} \right) \log \left( b_{\tau_2} + \frac{\nu}{2} \exp(v_{\tau_2}) \right),\\
     & \text{Leroux prior}: \quad \frac{\nu}{2} v_{\tau} - \left( \frac{\nu}{2} + a_{\tau} \right) \log \left( b_{\tau} + \frac{\nu}{2} \exp(v_{\tau}) \right) + \frac{1}{2}\log|G| + \frac{1}{2}v_{\rho} - \log (1+\exp(v_{\rho})).
\end{aligned}
\]
Using the analytical expressions of the hyperparameter distribution ensures fast inference of the model. Furthermore, by maximizing \eqref{eqn:log-posterior_hyper}, we obtain the maximum a posteriori estimates for $\boldsymbol{v_{\lambda}}$ and $ \boldsymbol{v_{\omega}}$, which are then used to derive the approximate conditional posterior of $\boldsymbol{\xi}$.

\section{Simulation study}

\subsection{Simulation set-up}
\begin{figure}[H]
     \centering
      \includegraphics[width=\textwidth]{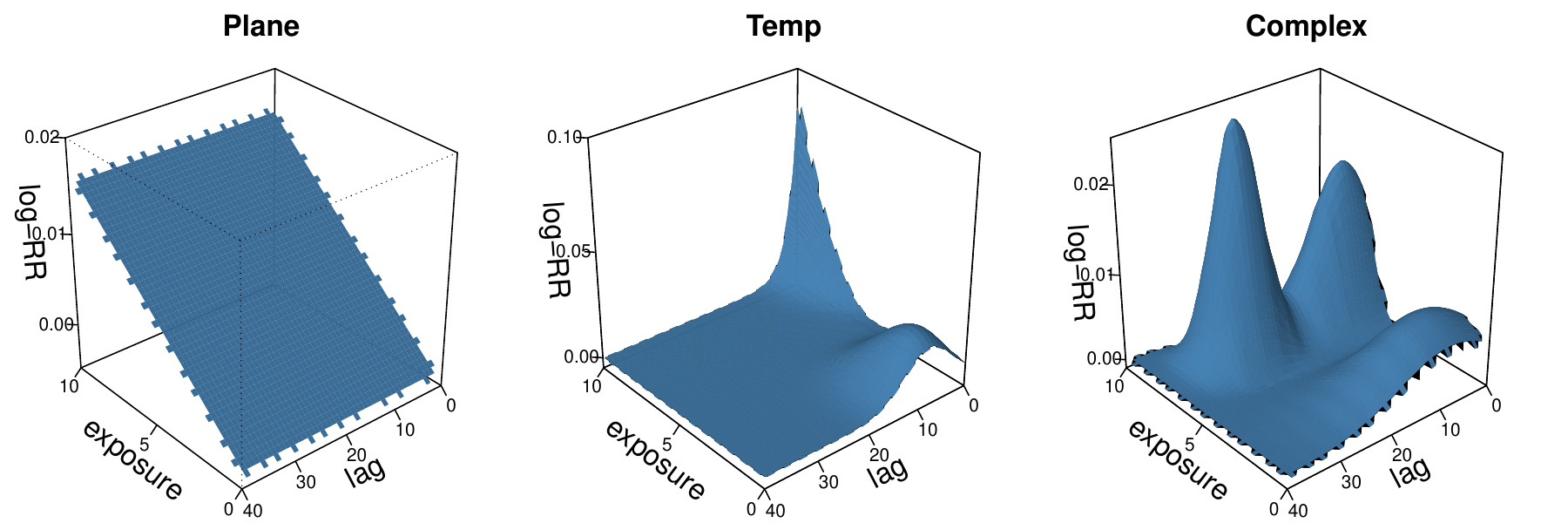}
      \caption{Simulation scenarios: (1) Plane, (2) Temp and (3) Complex. The log RR is presented as a function of the exposure variable and the lag.}
    \label{fig:all_scenarios}
\end{figure}

A simulation study is conducted to evaluate the performance of our proposed LPS method and to compare it with the standard fitting approach of \cite{gasparrini2017}. For the latter, the \texttt{bam} function in the R-package \texttt{mgcv} is used, which is a highly optimized way to fit a generalized additive model. It is designed for computational efficiency, especially for large datasets where the traditional \texttt{gam} function might be slow \citep{wood2015}. To optimize the smoothing parameters within this framework, REML is used as it is less prone to undersmoothing than alternative methods \citep{wood2011}. The performances of both methods are evaluated in the same three simulation scenarios that were considered by \cite{gasparrini2017}. Hence, the dose-lag-response function over lag $0-40$ can be described by: (1) a simple plane (`Plane'), (2) an expected temperature-mortality surface (`Temp') and (3) a more complex surface (`Complex'), as represented in Figure \ref{fig:all_scenarios}. More information about these specific scenarios can be found in \cite{gasparrini2017}. We consider these three scenarios in three situations: (1) one time series (`One'), (2) multiple time series with an independent random intercept for every area (`Indep.') and (3) multiple time series with an area-specific Leroux random effect (`Leroux'). 
In the first situation, we only have one time series containing $5114$ daily mortality counts simulated from a Poisson distribution in every scenario. The predictor is the daily temperature in Chicago, measured between 1987 and 2000 but standardized over the range $0-10$ \citep{samet2000}. In the other two situations, we make use of $983$ different time series containing $184$ observations each. The predictor in these situations is the daily temperature in the summer periods (June-August) of the years 2006 and 2013 in the $983$ different Middle Layer Super Output Areas (MSOA) in London. This predictor data was retrieved from \cite{gasparrini2022}. Like before, the predictor is standardized over the range $0-10$. A city-specific population size is simulated and included as an offset in the model. Mortality counts are generated from a Poisson distribution, where in addition to utilizing temperature as a predictor, an MSOA-specific random effect is incorporated into the model. This effect is assumed to be independent in Situation 2 but Leroux-structured with a $\rho$ parameter of $0.95$ in Situation 3.

P-splines with $10$ B-spline basis functions in both dimensions are used to fit the model, before imposing constraints. An additional quadratic varying ridge penalty is imposed, shrinking the effect to the null value at longer lags. The penalty matrix has the form $S_{l_2} = P_{v_l}+\delta I_{v_l}$ where $P_{v_l}$ is a diagonal matrix with weights given by $p_k = k^2$ with $k = 0, \ldots, (v_l-1)$. 

The model performance is evaluated using the coverage and root mean squared error (RMSE) of the lag-specific as well as overall (cumulative over a lag of $0-40$) log relative risk (RR), defined as
\begin{align*}
   \log\bigl(RR_{x,x_0}(l)\bigl) &=\sum_{i=1}^{v_x}\sum_{k=1}^{v_l}{\bigl(\widetilde{b}_i(x)-\widetilde{b}_i(x_0)\bigl)\shortsmile{b}_k(l)\theta_{ik}} 
\end{align*}
and
\begin{align*}
      \log\bigl(RR_{x,x_0}^{\text{overall}}\bigl) &=\sum_{i=1}^{v_x}\sum_{k=1}^{v_l}\biggl(\sum_{l=0}^{L}{\bigl(\widetilde{b}_i(x)-\widetilde{b}_i(x_0)\bigl)\shortsmile{b}_k(l)\biggl)\theta_{ik}} 
\end{align*}
respectively, comparing the risk at standardized exposure levels $x = 0,0.25, \ldots 9.75,10$ to a reference exposure level $x_0$.

Furthermore, the coverage and RMSE of the incidences and random effects are reported. Finally, we record the computation time (expressed in seconds) necessary for both methods. We perform $500$ simulation replicates in every scenario and each of the three situations. Due to large computation times, the simulations in Situation 2 (`Indep.') and Situation 3 (`Leroux') are performed using the VSC (Flemish Supercomputer Center).  

\subsection{Results of the simulation study}
Note that in Situation 1, the fitted model does not incorporate a spatial random effect as only one time series is assumed in this situation, while in Situation 2 an independent random effect is incorporated. In Situation 3, the LPS method incorporates a Leroux structure for the random effect, while \texttt{bam} fits only an independent random intercept as this package does not allow for the incorporation of a spatially structured random effect. The results of these simulations can be found in Table \ref{tab:results_simulation_update}, reporting the time expressed in seconds, RMSE and coverage. A figure showing the estimated and true lag-response curves at a scaled temperature of $8$ is included in the Supplementary Materials (Figure S1).


Examining the results for a single time series (`One'), Table \ref{tab:results_simulation_update} shows that both methods exhibit similar RMSEs and coverage values for the RR and incidence across Scenario 2 (`Temp') and 3 (`Complex'). Naturally, the RMSE and coverage of the random effect are not reported, as no random effect is included when analyzing a single time series. In the first scenario (`Plane'), where the relationship is linear and a high penalization is required, the \texttt{bam} method outperforms LPS in terms of RMSE. 
When introducing an independent random effect in the model, both methods continue to perform similarly in terms of RMSE in Scenario 2 (`Temp') and 3 (`Complex'), but the LPS method clearly provides superior (overall) RR coverage. In the first Scenario (`Plane'), \texttt{bam} still achieves a lower RMSE, yet the LPS method maintains good coverage and the RMSE remains relatively low. 
Similar observations can be made when incorporating a Leroux structured random effect, though an additional advantage emerges: the LPS method, assuming a Leroux structure, slightly reduces the RMSE of the random effect compared to \texttt{bam}, which is fitted under an independent structure assumption. This reduction also leads to slightly lower RMSEs for the incidence. Moreover, the assumption of a Leroux structure in LPS allows us to retrieve estimates of the underlying spatial process, such as the correlation parameter $\rho$. Estimating this correlation parameter is particularly useful in spatial analyses as it quantifies the strength of spatial dependence, helping to understand how strongly neighboring regions influence each other. The true assumed value of $\rho = 0.95$ is recovered fairly well by the LPS method, with a mean estimate of $0.92$ across simulations in all three scenarios. Also, the assumed spatial process variance of $0.5$ is accurately estimated, with a mean value of $0.46$ in each of the three scenarios.\\

Furthermore, computation time was evaluated using the \texttt{microbenchmark} package in R, with 10 function evaluations conducted for each simulation scenario with Leroux random effects. The analysis was done on a device with an Intel(R) Core(TM) i5-1135G7 CPU (base frequency 2.40GHz), having four cores and 16GB of RAM. As shown in Table \ref{tab:comptime}, the LPS method consistently demonstrates superior computational efficiency compared to the \texttt{bam} method. On average, LPS is approximately 4 to 6 times faster than \texttt{bam}, highlighting the computational benefit of our methodology.

\begin{table}[h]
        \centering
        \caption{Results of the simulations in the three situations (one time series (`One'), multiple time series with independent random effect (`Indep.') and multiple time series with Leroux random effect (`Leroux')). Results contain the RMSE of lag-specific log relative risk (RR), coverage of lag-specific relative risk, RMSE of overall log relative risk, coverage of overall relative risk, RMSE of random effect (RE), coverage of random effect, RMSE of incidence and coverage of incidence at 95\% nominal level in the three scenarios.}
        \scalebox{0.8}{
        \begin{tabular}{llcccccc}
        \hline
        \multicolumn{2}{c}{}& \multicolumn{2}{c}{One} & \multicolumn{2}{c}{\texttt{Indep.}} & \multicolumn{2}{c}{\texttt{Leroux}}\\
        \multicolumn{2}{c}{}& LPS & \texttt{bam} & LPS & \texttt{bam} & LPS & \texttt{bam} \\
        Scenario & Metric & \multicolumn{6}{c}{} \\
            \hline
        Plane & RMSE lag-specific RR & 0.0037 & 0.0011 & 0.0052 & 0.0022 & 0.0052 & 0.0028 \\
        & Coverage lag-specific RR & 0.9793 & 0.9307 & 0.9809 & 0.9483 & 0.9829 & 0.9330 \\
        & RMSE overall RR & 0.0716 & 0.0161 & 0.1885 & 0.0792 & 0.1906 & 0.0983\\
        & Coverage overall RR & 0.9009 & 0.9316 & 0.9468 & 0.9227 & 0.9505 & 0.9059 \\
        & RMSE RE & & & 0.2622 & 0.2616 & 0.1909 & 0.2256 \\
        & Coverage RE & & & 0.9407 & 0.9433 & 0.9698 & 0.9460 \\
        & RMSE incidence & 0.2723 & 0.1309 & 0.0040 & 0.0040 & 0.0029 & 0.0034\\
        & Coverage incidence & 0.9724 & 0.9316 & 0.9420 & 0.9437 & 0.9467 & 0.9444\\
        \hline 
        Temp & RMSE lag-specific RR & 0.0014 & 0.0013 & 0.0039 & 0.0034 & 0.0040 & 0.0035 \\
        & Coverage lag-specific RR & 0.9701 & 0.9529 & 0.9645 & 0.7435 & 0.9718 & 0.7417 \\
        & RMSE overall RR & 0.0163 & 0.0161 & 0.1068& 0.1109 & 0.1086 & 0.1143 \\
        & Coverage overall RR & 0.9643 & 0.9523 & 0.9859 & 0.5326 & 0.9924 & 0.5365\\
        & RMSE RE & & & 0.2893 & 0.2885 & 0.2064 & 0.2479\\
        & Coverage RE & & & 0.9357 & 0.9389 & 0.9667 & 0.9440\\
        & RMSE incidence & 0.9581 & 0.9547 & 0.0034 & 0.0034 & 0.0024 & 0.0028 \\
        & Coverage incidence & 0.9853 & 0.9116 & 0.9367 & 0.9389 & 0.9458 & 0.9421\\ 
        \hline
        Complex & RMSE lag-specific RR & 0.0027 & 0.0027 & 0.0038 & 0.0044 & 0.0039 & 0.0047 \\
        & Coverage lag-specific RR & 0.9750 & 0.9338 & 0.9667 & 0.8803 & 0.9672 & 0.8662 \\
        & RMSE overall RR & 0.0407 & 0.0412 & 0.1000 & 0.1314 & 0.1011 & 0.1427 \\
        & Coverage overall RR & 0.9261 & 0.9209 & 0.9355 & 0.7358 & 0.9298 & 0.7149\\
        & RMSE RE & & & 0.2778 & 0.2772 & 0.1996 & 0.2375\\
        & Coverage RE & & & 0.9423 & 0.9452 & 0.9682 & 0.9447\\
        & RMSE incidence & 0.2832 & 0.2942 & 0.0037 & 0.0037 & 0.0026 & 0.0030 \\
        & Coverage incidence & 0.9523 & 0.8994 & 0.9429 & 0.9447 & 0.9459 & 0.9428\\
        \hline
        \end{tabular}
        }
        \label{tab:results_simulation_update}
\end{table}

\begin{table}[h]
\centering
\caption{Comparison of computation time expressed in seconds with 10 evaluations between LPS with Leroux random effects and \texttt{bam} with independent random effects. The simulated data is generated with a Leroux spatial covariance.}
\begin{tabular}{llcccc}
\hline
\multicolumn{2}{l}{\textbf{}}  & \textbf{Minimum} & \textbf{Mean} & \textbf{Median} & \textbf{Maximum} \\ \hline
\multirow{2}{*}{Plane}   & LPS & 123.11           & 124.06        & 123.70          & 126.36           \\
                         & \texttt{bam} & 814.02           & 816.39        & 814.71          & 829.52           \\ \hline
\multirow{2}{*}{Temp}    & LPS & 195.59           & 196.46        & 196.50          & 197.56           \\
                         & bam & 925.57           & 928.12        & 926.86          & 939.29           \\ \hline
\multirow{2}{*}{Complex} & LPS & 135.76           & 136.22        & 136.26          & 136.75           \\
                         & bam & 751.73           & 752.54        & 752.55          & 753.68          \\ \hline
\end{tabular}
\label{tab:comptime}

\end{table}

\section{Data application}
We apply our method to a dataset containing the number of deaths in London, in the summer periods of the years 2006 and 2013 (June-August). The dataset is published by the Office of National Statistics (ONS) and has been used by \cite{gasparrini2022}. The daily total numbers of deaths in the age categories $0-74$ and $\geq 75$ are available across the $983$ MSOAs in London. The case counts of both age groups are summed to obtain the total number of daily deaths in every area. To adjust this number for the population size of each MSOA, we utilize population data from the year 2014 for all MSOAs, sourced from \cite{Londondatastore}. A figure showing the total number of deaths per $1000$ inhabitants during the summer periods of 2006 and 2013 in every MSOA can be found in the Supplementary Materials (Figure S2).  Daily temperature data can be retrieved from the HadUK-Grid product developed by the Met Office \citep{MetOffice2019}. Since this temperature data is available on a $1 \times 1$ km grid across the United Kingdom, an area-weighted average of the cells intersecting the MSOA is calculated to obtain an average temperature on the MSOA level, as done by \cite{gasparrini2022}. Spatial trends in temperatures can be seen (see Figure S3 in the Supplementary Materials).

\begin{figure}[H]
    \centering
    \includegraphics[width=0.90\linewidth]{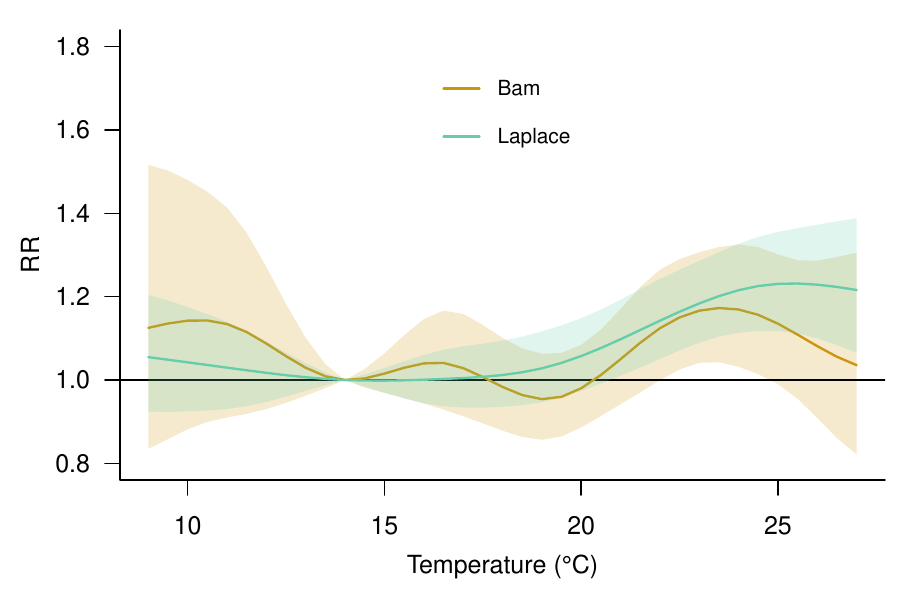}
    \caption{Exposure-response relationship represented by the overall relative risk, cumulated over 7 days. The blue colour represents the LPS method while the gold colour is the relationship fitted by the \texttt{bam} method. The lines represent point estimates, with shaded bands indicating 95\% credible/confidence intervals.}
    \label{fig:appl_leroux}
\end{figure}

\begin{figure}[H]
    \centering
    \includegraphics[width=0.85\linewidth]{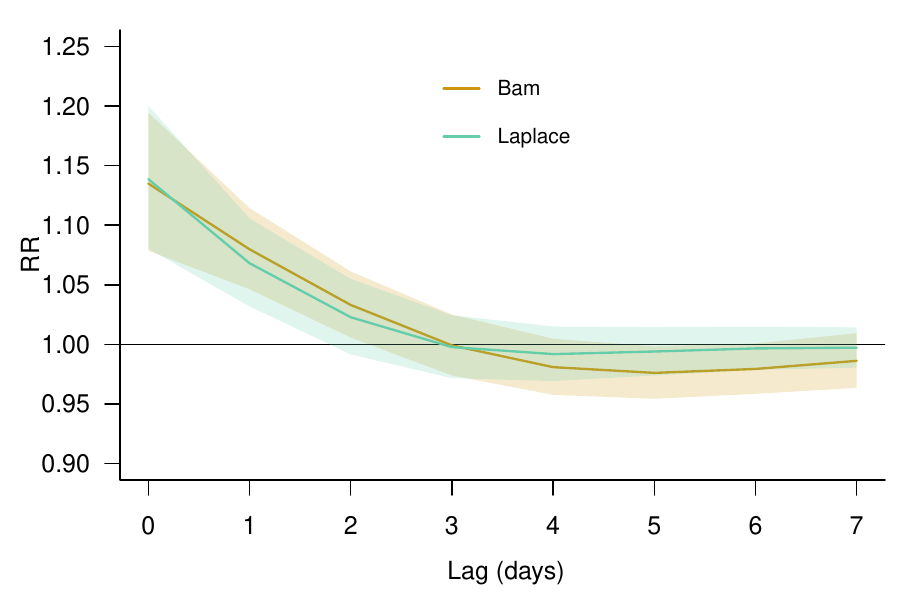}
    \caption{Lag-specific RR for a temperature of 24 degrees. The blue colour represents the LPS method while the gold colour is the relationship fitted by the \texttt{bam} method. The lines represent point estimates, with shaded bands indicating 95\% credible/confidence intervals.}
    \label{fig:lag_leroux}
\end{figure}

We specify a DLNM to analyze the relationship between the daily mortality counts, corrected for population size, and the average temperature. P-splines with $10$ B-spline basis functions in the variable as well as lag dimension are chosen, using a maximum lag of $7$ days. An additional penalty in the lag dimension is applied, defined identically to the one used in the simulation. Besides, we control for temporal variation by including an indicator variable for the day of the week as well as a natural cubic spline for the day of the year with $3$ degrees of freedom and an interaction with year. A Leroux random effect for every MSOA is assumed when using the LPS method, while an independent random effect is fitted with the \texttt{bam} method. Figure \ref{fig:appl_leroux} shows the overall relative risk (cumulated over $7$ days), compared to a reference temperature of $14^{\circ} \text{C}$, for the LPS method (blue) as well as the \texttt{bam} method (gold). We observe that the LPS method produces a smooth estimated curve, with a significantly increased risk for high temperatures, whereas the \texttt{bam} method seems to result in a somewhat overfitted curve. Moreover, due to the high uncertainty in the \texttt{bam}-estimated RR, no significant increase in overall RR is detected with the latter method. We can also look in more detail into the lag-specific RR for a scenario where the temperature is equal to 24 degrees Celsius, which is the $95$th temperature percentile. Figure \ref{fig:lag_leroux} shows a similar curve for both methods, with an increased RR for small lags. The LPS method appears to apply stronger shrinkage towards the null value at longer lags, compared to \texttt{bam}. The estimated random effects of both methods show similar overall patterns but differences can be observed between the estimates (see Figure S4 in the Supplementary Materials). The LPS method estimates the correlation parameter of the Leroux process as $\rho = 0.83$.\\

\begin{figure}[H]
    \centering
    \includegraphics[width=0.85\linewidth]{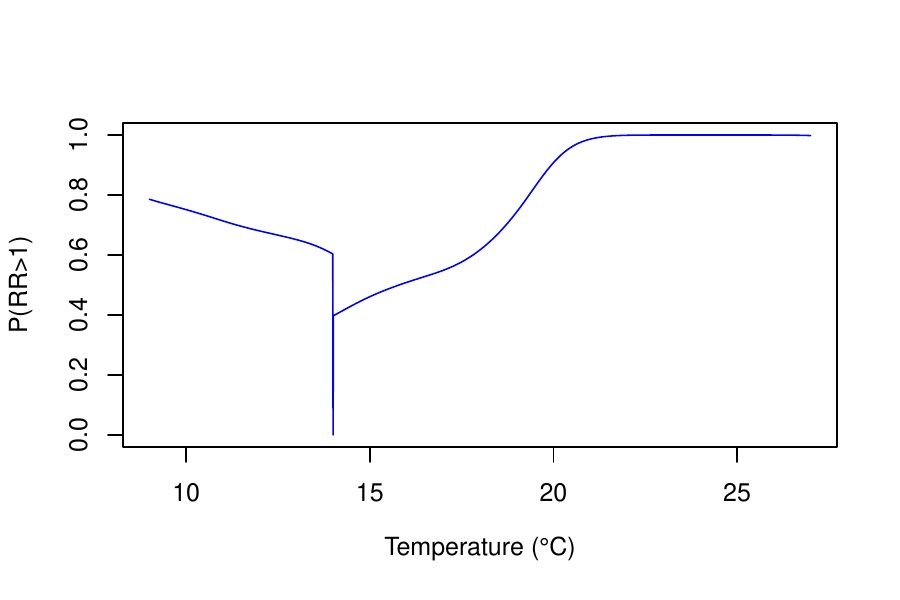}
    \caption{Probability that the overall RR exceeds 1 for different temperature values as compared to $14\,^\circ\mathrm{C}$.}
    \label{fig:exceedance_prob}
\end{figure}

\begin{figure}[H]
    \centering
    \includegraphics[width=\linewidth]{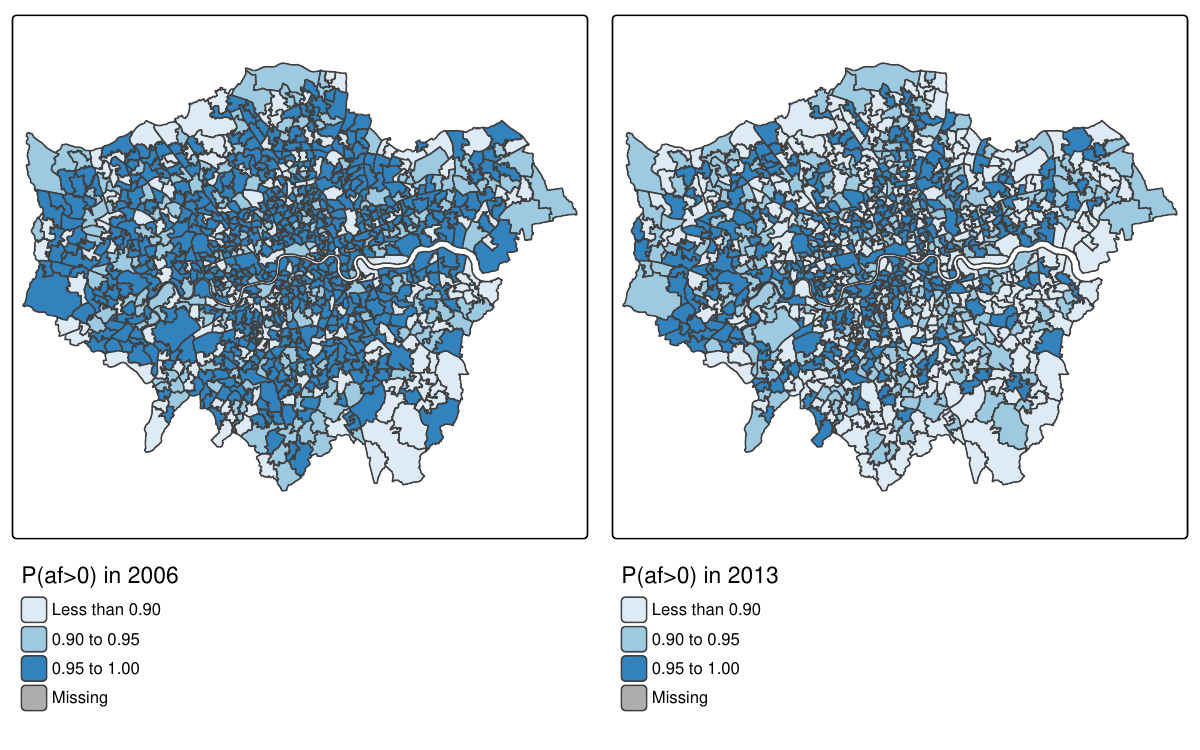}
    \caption{Probability that the AF exceeds 1 per MSOA.}
    \label{fig:exceedance_prob_af}
\end{figure}

A concept often of interest in a Bayesian analysis is the exceedance probability, i.e.,\ the probability that a particular quantity exceeds a particular threshold. Our LPS method makes these probabilities easy to calculate. For example, it might be worth exploring the probability that the overall RR for a specific temperature value exceeds 1. Figure \ref{fig:exceedance_prob} shows that for temperatures above $20\,^\circ\mathrm{C}$, this probability is above $95\%$. A sharp decline in the curve occurs at 14 degrees, which can be explained by this temperature being the reference value, resulting in an RR of exactly 1. Secondly, the LPS method allows us to directly capture the uncertainty in the attributable fraction, i.e.\ the fraction of deaths attributed to temperature, through the use of the posterior distribution. In every MSOA, we can therefore calculate the probability that the percentage of deaths in the year 2006 and 2013 attributable to temperature is strictly larger than zero, following the backward AF perspective. Figure \ref{fig:exceedance_prob_af} shows that in 2006, the exceedance probabilities were generally higher than in 2013, which is due to the larger average temperatures. We can also observe differences among the MSOAs, with some being more at risk than others.\\

We also compared our LPS method assuming a Leroux random effect with LPS methods assuming an ICAR, convolution or independent random effect. We found that the estimated cumulative relative risks from the three spatially structured models (i.e.,\ Leroux, ICAR and convolution) are (almost) indistinguishable. Very small differences could be found using the estimates obtained from a model assuming an independent random effect. These results are confirmed by the correlations between the estimated lag-specific relative risks, with a correlation of $1$ between the spatially structured models and a correlation of $0.998$ with the model with an independent random effect (see Figure S5 in the Supplementary Materials). To investigate the influence of the number of B-spline basis functions in the lag and variable dimension, we conducted a sensitivity analysis using (1) 6 B-spline basis functions in both dimensions and (2) 15 B-spline basis functions in both dimensions. The LPS method is seen to be more stable under these choices than the \texttt{bam} method (see Figure S6 in the Supplementary Materials). 

\section{Discussion and conclusion}

DLNMs are invaluable for studying the nonlinear lagged effects of exposure variables on outcomes of interest.  This paper proposes a Bayesian approach to DLNM with penalized splines to manage the flexibility of the fitted model. It also extends the penalized DLNM framework by incorporating random effects with CAR-type priors that account for spatial dependencies. Moreover, by leveraging the Laplace approximation for computational efficiency, the proposed Bayesian penalized DLNM offers significant advantages in handling large datasets.\\

Simulation studies show that the proposed Bayesian DLNM provides good coverage and low RMSE values overall. The \texttt{bam} routines and LPS perform similarly in terms of computational speed and accuracy when analyzing a single time series. However, with the inclusion of a random effect, the LPS method consistently demonstrates faster computation times as compared to \texttt{bam}. Additionally, in this context, LPS achieves better coverage of the relative risks in two out of three scenarios. Furthermore, when the random effect follows a Leroux structure, LPS outperforms \texttt{bam}, which assumes an independent random effect, in terms of RMSE for both the incidence and random effect. These results highlight the strengths of the LPS method, particularly in handling complex models with spatial random effects. The real data application further illustrates the practical utility of the proposed method. Applied to mortality data from London, the LPS method using spatially correlated random effects results in a smoother estimated overall relative risk curve compared to the \texttt{bam} function with an independent random effect. It is also worth noting that for this application, the LPS method with spatially structured random effects ($\approx$ 6 mins) is at least three times faster than the \texttt{bam} function with independent random effects ($\approx$ 20 mins).\\


In summary, our proposed approach offers a novel methodological contribution due to several advantages. First, the Bayesian framework enables better quantification of uncertainty, leading to improved coverage of relative risk estimates. Second, it facilitates ease of computation for quantities often of interest, such as exceedance probabilities and attributable fractions, due to the availability of posterior distributions. Third, the computational efficiency of the LPS method supports timely analysis, even for large datasets. Fourth, incorporating spatially structured random effects improves precision and further enhances coverage. Finally, to our knowledge, there is no existing methodology that integrates penalized DLNM with spatial random effects.\\

One of the limitations of our methodology is that it does not account for the uncertainty in the hyperparameters, specifically the penalty and spatial covariance parameters. We only use the a posteriori mode estimate for the hyperparameters to derive the conditional posterior of the latent parameters, similar to the empirical Bayes method. A potential solution to this is to use either a grid-based approach or the central composite design (CCD) for hyperparameter exploration \citep{rue2009}. However, these methods have the drawback of requiring additional computation time, especially in high-dimensional hyperparameter spaces. While CCD aims to enhance the computational efficiency compared to the grid-based strategy, it still poses significant computational demands. Alternatively, MCMC methods could be used to explore the posterior hyperparameter space. Although MCMC is computationally intensive and typically slower than sampling-free approaches, it offers an interesting alternative by allowing probabilistic exploration of hyperparameter distributions, which could provide valuable insights into parameter uncertainty. Another limitation of our study concerns the sensitivity of the chosen priors for the spatial variance parameters. While the chosen prior for the variance parameters has been shown to be robust for the penalty parameters \citep{jullion2007}, their sensitivity to the variance parameters remains to be investigated. Finally, exploring the use of R-INLA for the penalized DLNM would be interesting given its popularity and advanced development. However, it is important to note that implementing penalized DLNM in R-INLA is not straightforward, primarily due to the complex multidimensional structure of the penalty.\\

Finally, our model does not account for spatial variations in the exposure-lag relationship, unlike methods such as that of \cite{quijal2024}, which introduce spatially varying exposure-lag relationships within a non-penalized DLNM framework. Incorporating penalized DLNMs while allowing for spatially varying exposure-lag relationships at a fine geographical scale remains a complex challenge. Therefore, future work could also focus on developing a more flexible approach that integrates spatially varying risk within a penalized DLNM framework.

\section*{Supplementary materials}
\textbf{Online Appendix}: Contains additional figures for the data application. (Supplementary.pdf)\\


\section*{Acknowledgements}
The computational resources and services used in this work were provided by the VSC (Flemish Supercomputer Center), funded by the Research Foundation - Flanders (FWO) and the Flemish Government - department EWI.

\section*{Funding}

TN gratefully acknowledges funding by the Research Foundation - Flanders (grant number G0A3M24N).

\section*{Competing Interest Statement}
The authors have declared no competing interest.
\newpage
\bibliographystyle{apalike}
\bibliography{main}

\begin{thebibliography}{}

\bibitem[Analitis et~al., 2008]{analitis2008}
Analitis, A., Katsouyanni, K., Biggeri, A., Baccini, M., Forsberg, B., Bisanti, L., Kirchmayer, U., Ballester, F., Cadum, E., Goodman, P., et~al. (2008).
\newblock Effects of cold weather on mortality: results from 15 european cities within the phewe project.
\newblock {\em American Journal of Epidemiology}, 168(12):1397--1408.

\bibitem[Besag, 1974]{besag1974}
Besag, J. (1974).
\newblock Spatial interaction and the statistical analysis of lattice systems.
\newblock {\em Journal of the Royal Statistical Society: Series B (Methodological)}, 36(2):192--225.

\bibitem[Besag et~al., 1991]{besag1991}
Besag, J., York, J., and Molli{\'e}, A. (1991).
\newblock Bayesian image restoration, with two applications in spatial statistics.
\newblock {\em Annals of the Institute of Statistical Mathematics}, 43:1--20.

\bibitem[Chien et~al., 2018]{chien2018}
Chien, L.-C., Guo, Y., Li, X., and Yu, H.-L. (2018).
\newblock Considering spatial heterogeneity in the distributed lag non-linear model when analyzing spatiotemporal data.
\newblock {\em Journal of exposure science \& environmental epidemiology}, 28(1):13--20.

\bibitem[Eilers and Marx, 1996]{eilers1996}
Eilers, P. H.~C. and Marx, B.~D. (1996).
\newblock Flexible smoothing with {B}-splines and penalties.
\newblock {\em Statistical Science}, 11(2):89--121.

\bibitem[Eilers et~al., 2015]{eilers2015twenty}
Eilers, P. H.~C., Marx, B.~D., and Durb{\'a}n, M. (2015).
\newblock Twenty years of {P}-splines.
\newblock {\em SORT: Statistics and Operations Research Transactions}, 39(2):0149--186.

\bibitem[Fahrmeir and Lang, 2001]{fahrmeir2001}
Fahrmeir, L. and Lang, S. (2001).
\newblock Bayesian inference for generalized additive mixed models based on {M}arkov random field priors.
\newblock {\em Journal of the Royal Statistical Society Series C: Applied Statistics}, 50(2):201--220.

\bibitem[Gasparrini, 2022]{gasparrini2022}
Gasparrini, A. (2022).
\newblock A tutorial on the case time series design for small-area analysis.
\newblock {\em BMC Medical Research Methodology}, 22:129.

\bibitem[Gasparrini et~al., 2010]{gasparrini2010}
Gasparrini, A., Armstrong, B., and Kenward, M. (2010).
\newblock Distributed lag non-linear models.
\newblock {\em Statistics in Medicine}, 29(21):2224--2234.

\bibitem[Gasparrini and Leone, 2014]{Gasparrini2014}
Gasparrini, A. and Leone, M. (2014).
\newblock Attributable risk from distributed lag models.
\newblock {\em BMC Medical Research Methodology}, 14:55.

\bibitem[Gasparrini et~al., 2017]{gasparrini2017}
Gasparrini, A., Scheipl, F., Armstrong, B., and Kenward, M.~G. (2017).
\newblock A penalized framework for distributed lag non-linear models.
\newblock {\em Biometrics}, 73(3):938--948.

\bibitem[Gressani and Lambert, 2018]{gressani2018}
Gressani, O. and Lambert, P. (2018).
\newblock Fast {B}ayesian inference using {L}aplace approximations in a flexible promotion time cure model based on {P}-splines.
\newblock {\em Computational Statistics \& Data Analysis}, 124:151--167.

\bibitem[Gressani and Lambert, 2021]{gressani2021}
Gressani, O. and Lambert, P. (2021).
\newblock Laplace approximations for fast {B}ayesian inference in generalized additive models based on {P}-splines.
\newblock {\em Computational Statistics \& Data Analysis}, 154:107088.

\bibitem[Jullion and Lambert, 2007]{jullion2007}
Jullion, A. and Lambert, P. (2007).
\newblock Robust specification of the roughness penalty prior distribution in spatially adaptive {B}ayesian {P}-splines models.
\newblock {\em Computational Statistics \& Data Analysis}, 51(5):2542--2558.

\bibitem[Lang and Brezger, 2004]{lang2004}
Lang, S. and Brezger, A. (2004).
\newblock Bayesian {P}-splines.
\newblock {\em Journal of Computational and Graphical Statistics}, 13(1):183--212.

\bibitem[Lee, 2011]{lee2011}
Lee, D. (2011).
\newblock A comparison of conditional autoregressive models used in {B}ayesian disease mapping.
\newblock {\em Spatial and {S}patio-temporal {E}pidemiology}, 2(2):79--89.

\bibitem[Leroux et~al., 1999]{leroux1999}
Leroux, B., Lei, X., and Breslow, N. (1999).
\newblock Estimation of disease rates in small areas: A new mixed model for spatial dependence.
\newblock In Halloran, M.~E. and Berry, D., editors, {\em Statistical Models in Epidemiology, the Environment and Clinical Trials}, pages 135--178. Springer-Verlag, New York.

\bibitem[Lowe et~al., 2021]{lowe2021}
Lowe, R., Lee, S.~A., O'Reilly, K.~M., Brady, O.~J., Bastos, L., Carrasco-Escobar, G., de~Castro~Cat{\~a}o, R., Col{\'o}n-Gonz{\'a}lez, F.~J., Barcellos, C., Carvalho, M.~S., et~al. (2021).
\newblock Combined effects of hydrometeorological hazards and urbanisation on dengue risk in {B}razil: a spatiotemporal modelling study.
\newblock {\em The Lancet Planetary Health}, 5(4):e209--e219.

\bibitem[{Met Office} et~al., 2019]{MetOffice2019}
{Met Office}, Hollis, D., McCarthy, M., Kendon, M., Legg, T., and Simpson, I. (2019).
\newblock Hadukgrid gridded climate observations on a 1km grid over the uk, v1.2.0.ceda (1836-2022).
\newblock {\em Centre for Environmental Data Analysis}.

\bibitem[Montero et~al., 2010]{Montero2010}
Montero, J.~C., Mirón, I.~J., Criado-Álvarez, J.~J., Linares, C., and Díaz, J. (2010).
\newblock Mortality from cold waves in {Castile - La Mancha, Spain}.
\newblock {\em Science of the Total Environment}, 408:5768--5774.

\bibitem[{Office for National Statistics}, 2014]{Londondatastore}
{Office for National Statistics} (2014).
\newblock Super output area population {(LSOA, MSOA), London}: Land area and population density for {MSOA and LSOA} (2011) for most recent year.
\newblock \hypersetup{urlcolor=black}\url{https://data.london.gov.uk/dataset/super-output-area-population-lsoa-msoa-london}.
\newblock Last accessed: 15 July 2024.

\bibitem[Quijal-Zamorano et~al., 2024]{quijal2024}
Quijal-Zamorano, M., Martinez-Beneito, M.~A., Ballester, J., and Mar{\'\i}-Dell’Olmo, M. (2024).
\newblock Spatial {B}ayesian distributed lag non-linear models {(SB-DLNM)} for small-area exposure-lag-response epidemiological modelling.
\newblock {\em International journal of epidemiology}, 53(3):dyae061.

\bibitem[Rue et~al., 2009]{rue2009}
Rue, H., Martino, S., and Chopin, N. (2009).
\newblock Approximate {B}ayesian inference for latent {G}aussian models by using integrated nested {L}aplace approximations.
\newblock {\em Journal of the Royal Statistical Society: Series {B} (Statistical Methodology)}, 71(2):319--392.

\bibitem[Samet et~al., 2015]{samet2000}
Samet, J.~M., Zeger, S.~L., Dominici, F., Dockery, D., and Schwartz, J. (2015).
\newblock The national morbidity, mortality, and air pollution study (nmmaps). {P}art 1. methods and methodological issues. {T}echnical report.
\newblock {\em Health Effects Institute}.

\bibitem[Wood et~al., 2015]{wood2015}
Wood, S., Goude, Y., and Shaw, S. (2015).
\newblock Generalized additive models for large datasets.
\newblock {\em Journal of the Royal Statistical Society, Series C}, 64(1):139--155.

\bibitem[Wood, 2011]{wood2011}
Wood, S.~N. (2011).
\newblock Fast stable restricted maximum likelihood and marginal likelihood estimation of semiparametric generalized linear models.
\newblock {\em Journal of the Royal Statistical Society. Series B (Statistical Methodology)}, 73(1):3--36.

\end{thebibliography}

\end{document}


\maketitle

\newpage
\subsection*{Web Figures}

\begin{figure}
    \centering
    \includegraphics[width=0.85\linewidth]{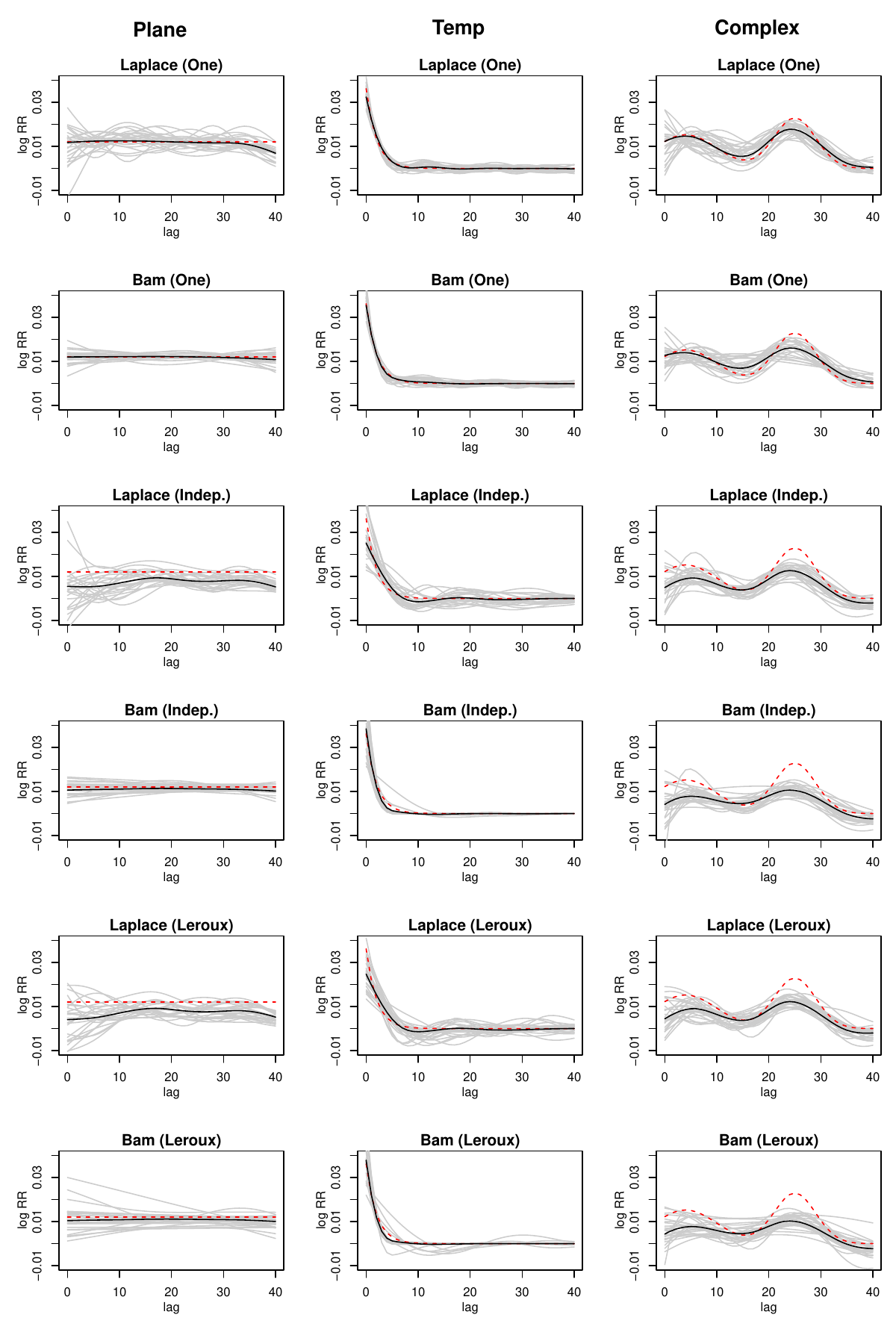}
    \caption{Results of the simulation study, illustrating the performance of LPS and \texttt{bam}. The figures represent the lag-response curve for the situation with one time series (One) in row 1-2, situation with independent random effect (`Indep.') in row 3-4 and situation with a Leroux random effect (`Leroux') in row 5-6. The bold black line represents the mean estimate across 1000 replicates while the dotted red line represents the true log RR. The grey curves are the fits from 25 replicates.}
    \label{fig:results_simulation}
\end{figure}

\begin{figure}[h]
    \centering
    \includegraphics[width=0.90\linewidth]{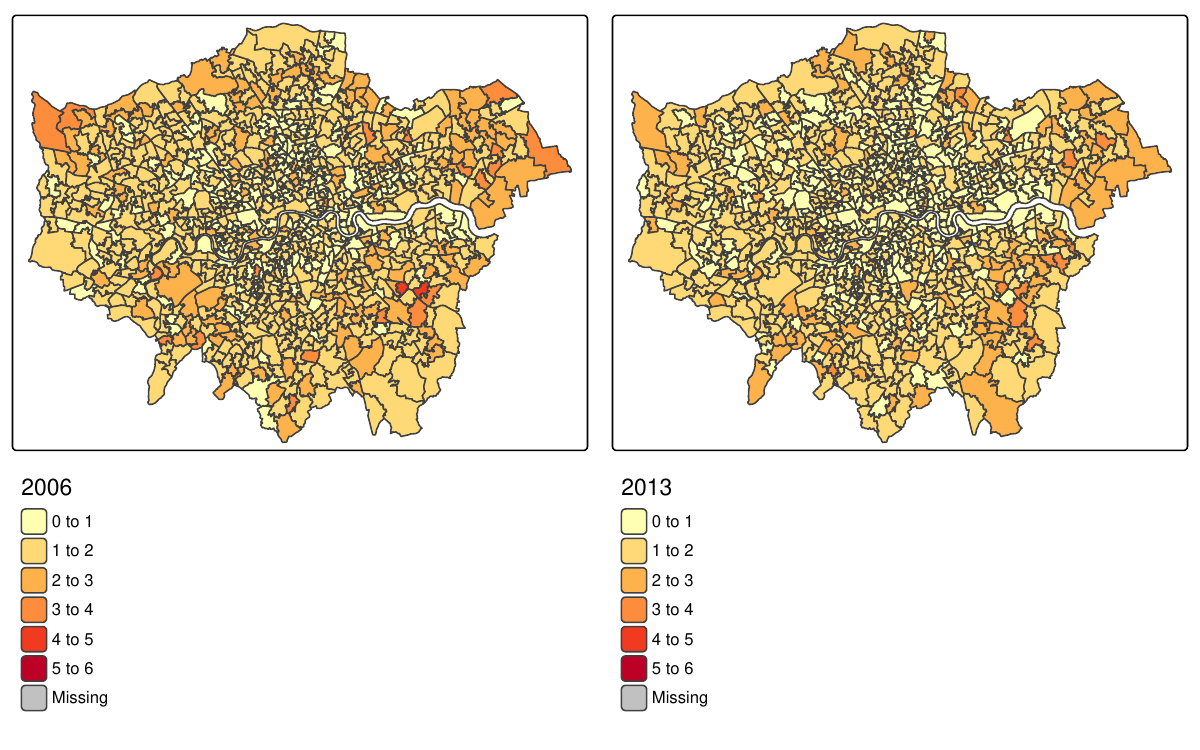}
    \caption{Total number of deaths per $1000$ inhabitants during the summer period of 2006 (left) and 2013 (right) in every MSOA.}
    \label{fig:map_deaths}
\end{figure}

\begin{figure}[h]
    \centering
    \includegraphics[width=0.90\linewidth]{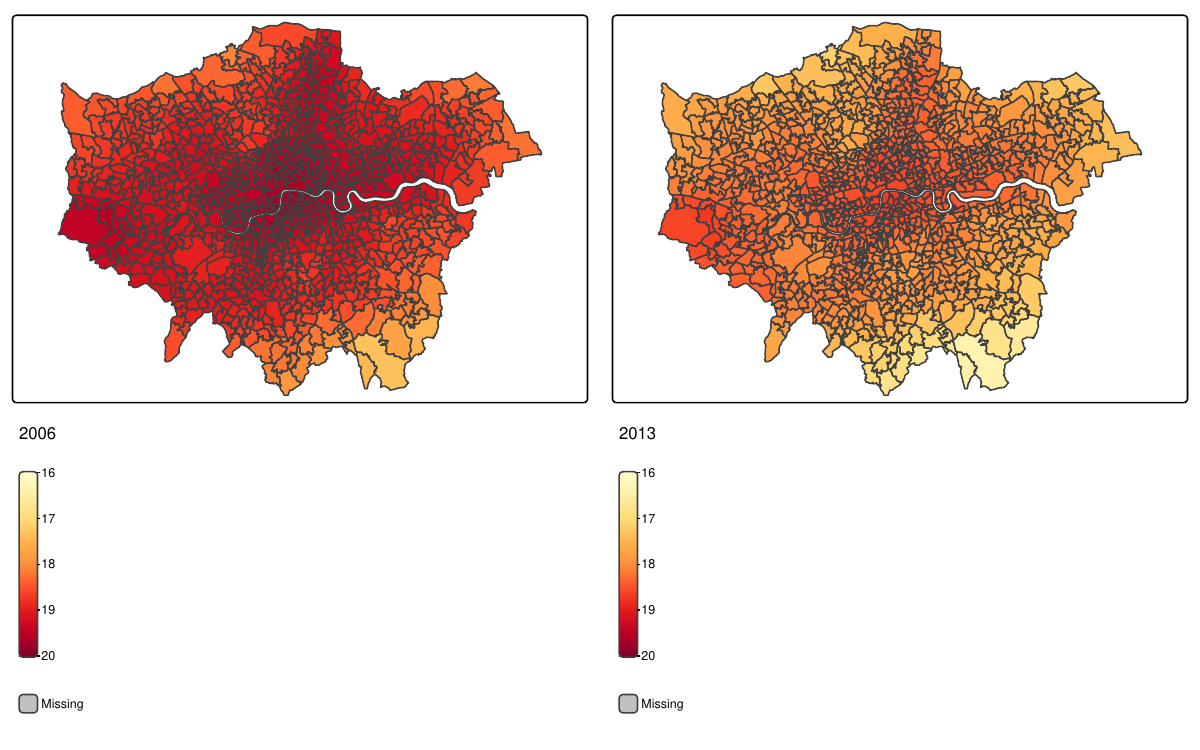}
    \caption{Average temperature during the summer period of 2006 (left) and 2013 (right) in every MSOA.}
    \label{fig:map_temp}
\end{figure}

\begin{figure}[h]
    \centering
    \includegraphics[width=0.90\linewidth]{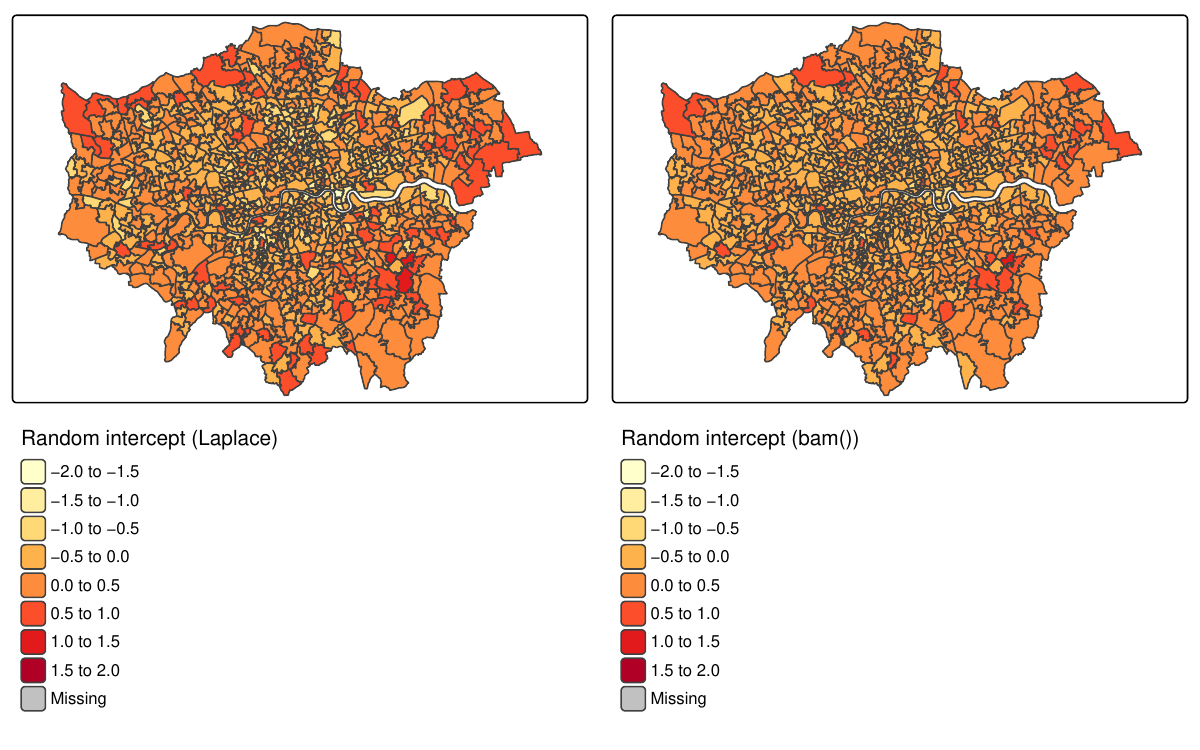}
    \caption{Random effect estimated by LPS and \texttt{bam} method.}
    \label{fig:random_Leroux}
\end{figure}

\begin{figure}[h]
    \centering
    \includegraphics[width=\linewidth]{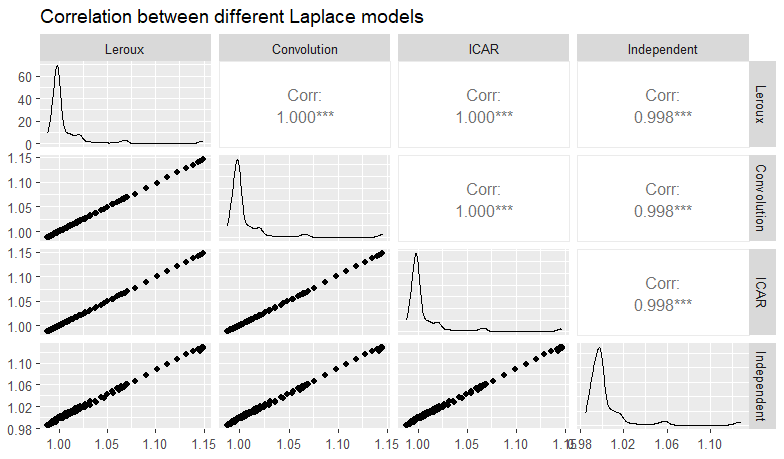}
    \caption{A plot showing the pairwise correlations between the lag-specific estimated relative risks of a LPS model with Leroux, Convolution, ICAR and independent random effects.}
    \label{fig:correlation_laplace}
\end{figure}

\begin{figure}[h]
    \centering
    \includegraphics[width=\linewidth]{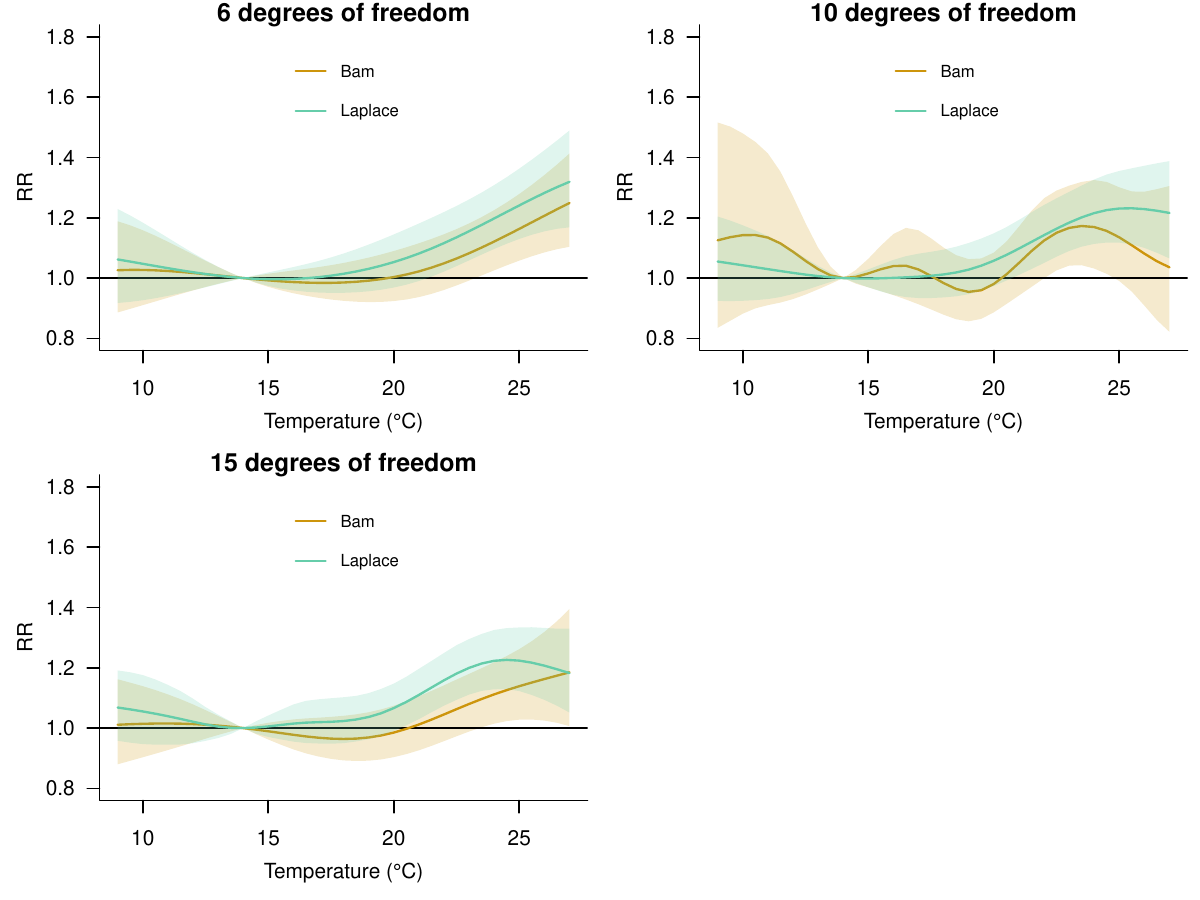}
    \caption{Sensitivity analysis: Exposure-Response relationship represented by the overall RR, over 7 days for splines with (1) 6 degrees of freedom, (2) 10 degrees of freedom and (3) 15 degrees of freedom.}
    \label{fig:sensitivity_analysis}
\end{figure}